\documentclass[10pt, conference, letterpaper]{IEEEtran}
\IEEEoverridecommandlockouts

\usepackage{cite}
\usepackage{amsmath,amssymb,amsfonts,bm}
\usepackage{algorithmic}
\usepackage{graphicx}
\usepackage{textcomp}
\usepackage{xcolor}
\usepackage{subfigure}
\usepackage{tikz}
\usepackage{xspace}
\usepackage[belowskip=-15pt,aboveskip=0pt,font=footnotesize]{caption}
\def\BibTeX{{\rm B\kern-.05em{\sc i\kern-.025em b}\kern-.08em
    T\kern-.1667em\lower.7ex\hbox{E}\kern-.125emX}}
\begin{document}

\title{ProSpire: \textbf{Pro}active \textbf{S}patial \textbf{P}red\textbf{i}ction of \textbf{R}adio \textbf{E}nvironment Using Deep Learning
}

\author{
\IEEEauthorblockN{
    Shamik Sarkar\textsuperscript{1},
    Dongning Guo\textsuperscript{2},
    Danijela Cabric\textsuperscript{1}
}
\IEEEauthorblockA{
    \textsuperscript{1}University of California, Los Angeles,
    \textsuperscript{2}Northwestern University 
}
\IEEEauthorblockA{
    \textsuperscript{}shamiksarkar@ucla.edu,
    \textsuperscript{}dGuo@northwestern.edu 
    \textsuperscript{}danijela@ee.ucla.edu 
}
}

\newcommand{\sys}{ProSpire\xspace}
\newcommand{\DL}{DL\xspace}
\newcommand{\NN}{NN\xspace}

\maketitle
\begin{abstract}
Spatial prediction of the 
radio propagation environment (henceforth `radio environment' for brevity) of a transmitter can assist and improve various aspects of wireless networks. The majority of research in this domain can be categorized as `reactive' spatial prediction, where the predictions are made based on a small set of measurements from an active transmitter whose radio environment is to be predicted. Emerging spectrum-sharing paradigms would benefit from `proactive' spatial prediction of the radio environment, where the spatial predictions must be done for a transmitter for which no measurement has been collected. 

This paper proposes a novel, supervised deep learning-based framework, \sys, that enables spectrum sharing by leveraging the idea of proactive spatial prediction. We carefully address several challenges in \sys, such as designing a framework that conveniently collects training data for learning, performing the predictions in a fast manner, enabling operations without an area map, and ensuring that the predictions do not lead to undesired interference. \sys relies on the crowdsourcing of transmitters and receivers during their normal operations to address some of the aforementioned challenges. The core component of \sys is a deep learning-based image-to-image translation method, which we call RSSu-net. We generate several diverse datasets using ray tracing software and numerically evaluate \sys. Our evaluations show that RSSu-net performs reasonably well in terms of signal strength prediction, $\approx$ 5 dB mean absolute error, which is comparable to the average error of other relevant methods. Importantly, due to the merits of RSSu-net, \sys creates proactive boundaries around transmitters such that they can be activated with $\approx$ 97\% probability of not causing interference. In this regard, the performance of RSSu-net is 19\% better than that of other comparable methods.
\end{abstract}


\section{Introduction} \label{section:intro}
The problem of spatial prediction of the radio environment is to predict the received signal strength (RSS), from a transmitter, at various locations where no receivers are present. There are many applications of spatial prediction of the radio environment, e.g., path planning~\cite{krijestorac2021spatial}, 
spectrum sharing~\cite{achtzehn2012improving}, etc. 
Most research in this domain can be categorized as `reactive' spatial prediction, where the predictions are made using a small set of RSS measurements from an active transmitter whose radio environment is to be predicted. However, emerging spectrum-sharing paradigms, e.g., national radio dynamic zones (NRDZ), can benefit from `proactive' spatial prediction of the radio environment, where the predictions must be made for transmitters for which no measurement is available. 

\textbf{Limitations of reactive spatial prediction}: In shared-spectrum wireless networks, reactive spatial prediction has several limitations. First, the collected RSS measurements can have errors. Consider an example where we want to predict the radio environment of transmitter $T_1$. For reactive spatial prediction, first, a set of receivers must collect a small set of RSS measurements from $T_1$. However, the receivers' measurements can be erroneous as they may be interfered by another coexisting transmitter, say $T_2$, $T_2 \neq T_1$, as the spectrum is shared. Second, the need for a preliminary set of RSS measurements can cause interference to the receivers of a coexisting shared-spectrum network. Again using the above example, if we want to predict the radio environment of $T_1$, then $T_1$ will have to transmit, and a small set of RSS measurements must be collected. However, any transmission from $T_1$ may cause interference to the receivers of $T_2$. Third, in certain scenarios, it may be required to use a dedicated set of receivers just to collect the small set of RSS measurements that are mandatory for reactive spatial prediction~\cite{maeng2022out}. Fourth, collecting a small set of RSS measurements from an active transmitter implies utilizing resources, e.g., power, bandwidth, and time, just to make the reactive spatial predictions.

\textbf{Goal of our work}: 
In this work, we investigate the potential of proactive spatial prediction in shared-spectrum wireless networks. Proactive spatial prediction does not suffer from the limitations of reactive spatial prediction as it does not require any RSS measurements from the transmitter whose radio environment is to be predicted. We consider a general spectrum-sharing setup where the spectrum is primarily dedicated to one wireless network (primary), and transmitters of another network (secondary) opportunistically use the same spectrum without causing harmful interference to the receivers of the primary network (henceforth primary receivers). 
An example of such a spectrum-sharing setup is NRDZ~\cite{zheleva2023radio}. The idea of NRDZ is to create dynamic radio zones around experimental wireless transmitters such that they can be operated without interfering with the existing communication system receivers in the vicinity. 
\emph{Our goal is to devise an efficient framework for proactive spatial prediction for transmitters of the secondary network} (henceforth secondary transmitters). These predictions will guide the secondary transmitters to actively transmit such that the interference to the primary receivers is tolerable. 
Here `efficient' implies 
not requiring additional infrastructure deployment just to perform the proactive spatial predictions.

\textbf{Challenges in proactive spatial prediction}: While  proactive spatial prediction is advantageous over its reactive counterpart, 
the goal of 
proactive spatial prediction 
for spectrum sharing 
brings its own challenges.
\begin{itemize}
    \item First, if no measurements are available for a transmitter, what should be the basis for predicting its RSS? A simple approach is to use the radio wave propagation path loss model (PLM)~\cite{rappaport1996wireless} that relates RSS to the distance between a transmitter and the target location where RSS is to be predicted. However, 
    such an approach 
    does not take into account the blockages in the environment.
    \item Second, the proactive spatial predictions must be made quickly; otherwise 
    the temporal transmission opportunities in a shared spectrum network can be limited. For example, proactive spatial prediction can be made via ray tracing software, but that is very time-consuming. 
    \item Third, proactive spatial prediction must be feasible even when detailed maps are unavailable. 
    This challenge stems from the fact that spectrum sharing may be needed in any geographical area, and assuming the availability of detailed maps for all such areas is unrealistic.
    This is another reason for ray tracing not being applicable.
\end{itemize}

\textbf{Overview of \sys}: In this paper, we present a general learning-based framework, \sys, that enables spectrum sharing among the primary and secondary networks by leveraging the idea of proactive spatial prediction of the radio environment of transmitters. \sys carefully addresses all the challenges related to proactive spatial prediction.

To address the first challenge, \sys relies on supervised learning, specifically deep learning (DL), as the basis for its predictions. We propose a crowdsourcing approach for gathering the training data that will be used for learning in \sys. In our crowdsourcing model, the transmitters and receivers of the primary network report their measurements and activities during normal operations. The crowdsourced data is gathered by a centralized spectrum administrator. Using the collected data, \sys learns a generalized model that can accurately capture the radio environment. Our intuition is that if the training data is diverse and the learning is generalized, we can make RSS predictions for transmitter-receiver location pairs for which we have no measurement in the training data.

As \sys uses \DL-based predictions, it is inherently fast. Thus, \sys can address the second challenge of proactive spatial prediction. For deep neural network (NN) based predictions, we develop an approach called RSSu-net which uses an image-to-image (I2I) translation method. RSSu-net is built upon the u-net \NN architecture. For a given transmitter location, if the RSS values are to be predicted at multiple locations, RSSu-net has the advantage of being fast as it can make all the predictions by a single pass of a \NN. 

One way to address the third challenge is not to use maps at all. However, recent works have shown the benefits of using maps in \DL-based spatial prediction~\cite{krijestorac2021spatial}. Hence, to exploit the benefits of maps while addressing the third challenge, we propose to use radio tomographic imaging (RTI) based maps instead of detailed maps in our \DL approach. Our RTI-based \DL is robust to obstacles not present on maps, e.g., foliage. 

After proactive RSS prediction for the secondary transmitters, the next task is to use the predictions for activating the secondary transmitters such that the interference to the primary receivers is tolerable. For this, we develop a boundary proposal algorithm that operates on the proactive spatial predictions for a secondary transmitter. This algorithm proposes a zone around the transmitter and a power level, $z_{ooz}$, such that the RSS out of the zone will be below $z_{ooz}$ dBm if the transmitter becomes active at full power. Based on this zone, we adapt the secondary transmitter's power before activating it such that the chances of interference to primary receivers are reduced. Note that the performance of boundary proposal and power adaptation depends on the accuracy of the proactive predictions. If the predicted RSS values are much higher than actual RSS values (i.e., overestimation), the utilization of the shared spectrum can be poor. On the other hand, if the predicted RSS values are much lower than actual RSS values (i.e., underestimation), transmissions based on the predictions can cause high interference to the primary receivers. With this in mind, we use a loss function in RSSu-net that penalizes underestimations more than overestimations 
while maintaining a low 
average estimation error. I.e., we find a balance between the two types of errors while prioritizing interference protection to primary receivers over transmission opportunities of the secondary transmitters.

\textbf{Contributions}: In summary, our main contributions are:
\begin{itemize}
    \item We propose a novel framework called \sys that enables spectrum sharing by leveraging the idea of proactive spatial prediction. \sys relies on crowdsourcing measurements from the primary transmitters and receivers.   
    \item We develop a \DL approach, RSSu-net, that uses the crowdsourced measurements and learns to perform proactive spatial prediction for the secondary transmitters.
    \item We incorporate RTI in \sys to deal with the unavailability of maps and also assist our DL model.
    \item We develop a boundary proposal algorithm that activates secondary transmitters in a non-interfering manner.    
\end{itemize}

\section{Related Work}\label{section:related_work}
\textbf{Reactive spatial prediction}: Reactive spatial prediction relies on signal processing methods like Kriging interpolation~\cite{maeng2022out}, matrix completion~\cite{chouvardas2016method}, tensor completion~\cite{schaufele2019tensor},
etc. Recently, \DL has also been applied to reactive spatial prediction using u-net~\cite{levie2021radiounet}, autoencoders~\cite{teganya2021deep}, and ResNets~\cite{li2021model}. These works have shown that \DL-based reactive spatial prediction methods perform better than signal-processing methods. Broadly, all these methods are interpolation methods that rely on sparse measurements to construct a dense representation of the radio environment.
A detailed survey of reactive spatial prediction methods can be found in~\cite{romero2022radio}. 

\textbf{Proactive spatial prediction}: If the predictions are in the form of RSS values, the simplest approach for proactive spatial prediction is to use the PLM~\cite{rappaport1996wireless}. The idea of proactive spatial prediction can be used for channel prediction. For example, a \DL model can predict the channel on device-to-device links based on measurements from the cellular channel~\cite{najla2020predicting}. Another way of performing proactive spatial prediction is to collect training data from different cities, learn a \DL model, and perform predictions on an entirely new area~\cite{levie2021radiounet}. However, this approach cannot be used in \sys as it relies heavily on city maps. Finally, proactive spatial prediction can be performed using ray tracing software, but that is time-consuming and will not work if the map of the area is unavailable.

\section{\sys Framework} 
\label{section:system_model}
We consider an $(Ll)$ m $\times$ $(Wl)$ m area where $L$, $W$, and $l$ are integers. We define the area length and width as multiples of $l$ for subsequent notational convenience. This area has $T_{PN}$ static primary transmitters, each serving several mobile users. We denote $\mathcal{T}$ as the set of primary transmitters' locations. The primary transmitters can be cellular base stations or WiFi access points, but we do not impose any such restrictions. The primary transmitters operate on the same frequency band of $B$ MHz and use an omnidirectional antenna with a fixed transmit power of $P_{PN}$ dBm. The primary transmitters do not interfere with each other and coordinate their transmissions via TDMA or CSMA. We assume all the primary transmitters and receivers are located outdoors.

\begin{figure}[t]
    \centering
    \begin{subfigure}[Operations during data collection phase]
{\label{fig:system_model_training}
    \includegraphics[scale=0.295]{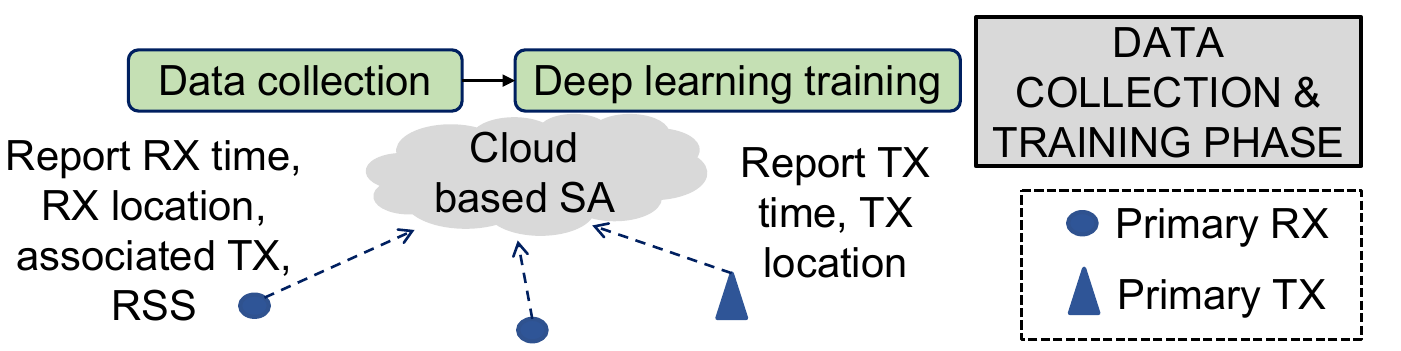}}
    \end{subfigure}
    \begin{subfigure}[Operations in the online phase]
{\label{fig:system_model_testing}
    \includegraphics[scale=0.295]{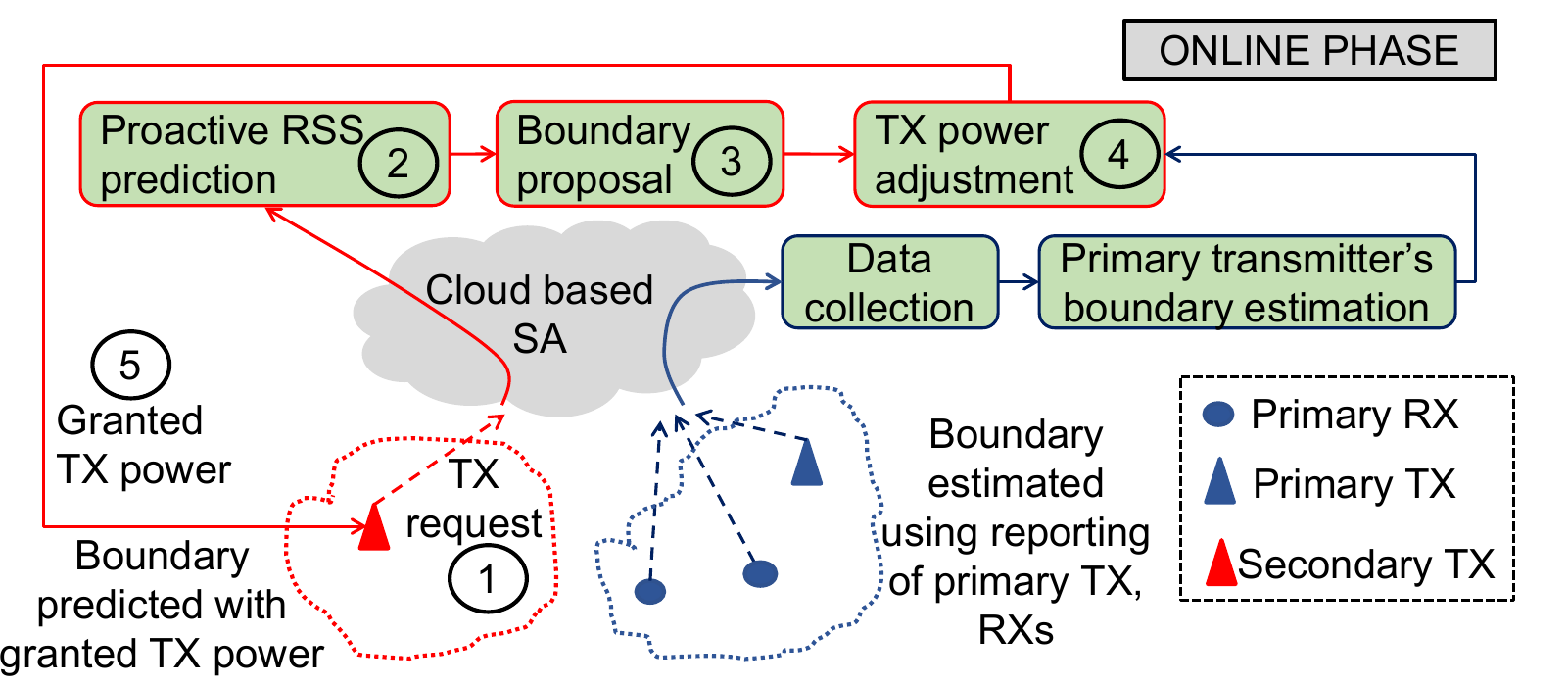}}
    \end{subfigure}
    \vspace{-1mm}
    \caption{Figure shows the operations of \sys.
    }
    \label{fig:system_model}
\vspace{-1mm}
\end{figure}

\textit{Data Collection:} We consider a cloud-based spectrum administrator (SA), as shown in Fig.~\ref{fig:system_model_training}, which crowdsources measurements from the primary transmitters and receivers. Specifically, the primary transmitters report their transmit time and location, whereas the primary receivers report their measurement time, location, associated transmitter, and RSS (averaged over small-scale fading) measurements as shown in Fig.~\ref{fig:system_model_training}. Using the collected data, the SA creates a dataset, 
$\mathcal{D} = \big\{(u,v, \{(x, y, z); (x, y) \in \mathcal{R}_{(u,v)}\}\big); (u,v) \in \mathcal{T}$\}
where $(u,v)$ is a primary transmitter's location, $(x, y)$ is a primary receiver's location, and $z$ (dBm) is its measured RSS from the transmitter at $(u,v)$. $\mathcal{R}_{(u,v)}$ is a set of receivers that measured RSS from the transmitter at $(u,v)$. All the measurements of a particular entry in $\mathcal{D}$ (i.e., for a particular primary transmitter since $|\mathcal{D}| = T_{PN}$) need not be collected simultaneously. This way, the SA can collect measurements for every primary transmitter at many locations. Finally, using $\mathcal{D}$, the SA trains a \DL model for proactive spatial predictions in the online phase.

\textit{Online Phase:} In this phase, the SA provides opportunities for the secondary transmitters to use the same spectrum on which the primary transmitters operate. First, a secondary transmitter located at $(u',v')$ requests permission from the SA to use the spectrum; see \tikz \node[draw,circle, inner sep=-1pt, minimum size=2.5mm]{\footnotesize 1}; in Fig.~\ref{fig:system_model_testing}. 
Based on $(u',v')$, the SA performs proactive spatial prediction of the RSS values for the secondary transmitter at several locations around it; see \tikz\node[draw,circle,inner sep=-3pt, minimum size=2.5mm]{\footnotesize 2}; in Fig.~\ref{fig:system_model_testing}. Next, among the locations where RSS has been predicted, the SA selects a set of locations using the boundary proposal algorithm, which defines the proposed boundary for the secondary transmitter; see \tikz \node[draw,circle,inner sep=-3pt, minimum size=2.5mm]{\footnotesize 3}; in Fig.~\ref{fig:system_model_testing}. The proposed boundary is a zone beyond which the RSS due to the secondary transmitter would be below
$z_{ooz}$ dBm, if the transmitter becomes active with power $P_{PN}$ dBm. We call $z_{ooz}$ out-of-zone power leakage. Note that this boundary proposal must be made before activating the secondary transmitter. 

The proactive spatial prediction
method assumes that the secondary transmitter will use the same power and antenna characteristics as the primary transmitters.
This assumption is needed for the functioning of the \DL model, which expects the training and test data to be generated by the same distribution.
However, there can be a set of active primary transmitters with an interference protection requirement. E.g., a primary transmitter can specify that the interference at any of its receivers should not exceed $\alpha$ dB above the noise floor ($N_f$ dBm). Using the locations of the active primary transmitters and their specified interference protection, the SA can estimate an interference protection boundary around each of them. One such interference protection boundary is shown in Fig.~\ref{fig:system_model_testing} using blue dots. Note that, unlike the boundary proposal for the secondary transmitter, the interference protection
boundary estimation for active primary transmitters is not proactive. Thus, the interference protection boundary for the primary transmitters can be estimated using a small set of measurements from the primary receivers and any of the reactive spatial prediction methods discussed in Section~\ref{section:related_work}. Based on the interference protection boundary of the active primary transmitters, the SA adjusts the transmit power of the secondary transmitter to $P_{SN}$ $(\leq P_{PN})$ such that there is no interference to the primary receivers; see \tikz \node[draw,circle,inner sep=-3pt, minimum size=2.5mm]{\footnotesize 4}; in Fig.~\ref{fig:system_model_testing}. 
Finally, the SA responds to the secondary transmitter with the permission to transmit at the power level of $P_{SN}$ dBm; see \tikz \node[draw,circle,inner sep=-3pt, minimum size=2.5mm]{\footnotesize 5}; in Fig.~\ref{fig:system_model_testing}. Fig.~\ref{fig:system_model_testing} shows a red dotted boundary around the secondary transmitter that represents the proactively proposed boundary for that transmitter at granted power of $P_{SN}$ dBm. 

\section{Building blocks of \sys} \label{section:methods}
In this section, we describe the building blocks of \sys.

\subsection{Proactive Spatial Prediction} \label{subsection:unet}
Our goal is to have a method for proactive spatial prediction of the radio environment for secondary transmitters without any active transmission from them. Specifically, using $\mathcal{D}$, we have to learn a function $\mathbf{f}$ that maps a secondary transmitter's location to the RSS values (had the secondary transmitter been active with $P_{PN}$ dBm transmit power) at a set of queried locations in the transmitter's vicinity. An important question here is how to define the set of query locations. To answer that, we first divide the whole area in a rectangular grid such that each grid voxel is of size $l$ m $\times$ $l$ m. In the online phase, for a given secondary transmitter at $(u',v')$, $(u',v') \notin \mathcal{T}$, we define the set of query points, $\mathcal{K}$; $|\mathcal{K}| = K$, as grid points that are nearest to $(u',v')$ but are not under buildings or foliage in that area. We use the term grid point to imply the center of a grid voxel. We choose the query points this way because we primarily care about predicting RSS at locations near the secondary transmitter. Locations far away from the transmitter would have low values of RSS, which implies a lower possibility of causing interference to the primary receivers at those far away locations.

Now, we describe our algorithm for proactive prediction of RSS at $K$ nearest grid points for a given secondary transmitter location. For this, we use a \DL-based I2I approach. We use an I2I approach because of the following reason. I2I methods are usually built using convolutional \NN (CNN) that exploit the spatial correlation of the pixels in the input image.  In our problem, we frame the input to our \NN as a set of images with a certain spatial structure (details in the following paragraph), and CNN-based I2I models can efficiently learn from such structure. Additionally, it has the advantage of being fast as it can make all the $K$ predictions by a single pass of the trained \NN, $\mathbf{f}$. In the following, we explain our training method followed by the inference procedure.
\begin{figure}[t]
    \centering
    \includegraphics[scale=0.43]{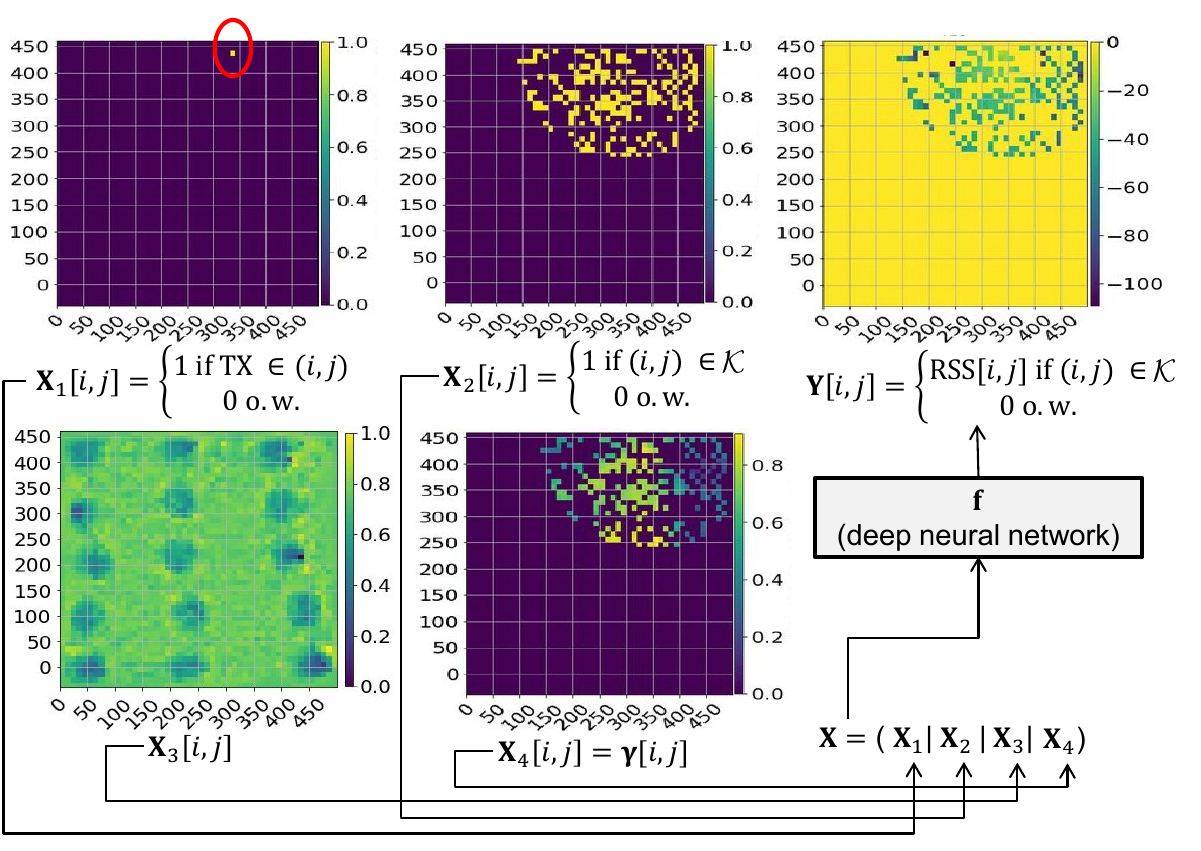}
    \caption{Figure shows the inputs and output of the deep \NN in \sys. 
    }
    \label{fig:I2I_method}
\end{figure}

\textit{Training}: Our training procedure for learning $\mathbf{f}$, which is represented by a \NN, is based on the collected dataset, $\mathcal{D}$. Recall from Section~\ref{section:system_model}, that $|\mathcal{D}| = T_{PN}$. Each of the entries in $\mathcal{D}$ becomes a training example in our training dataset. Hence, each of our training examples corresponds to the transmission of a primary transmitter and the associated RSS measurement at different primary receiver locations. For a training example, the input to $\mathbf{f}$ is a tensor or 3-D volume 
$\mathbf{X} = (\mathbf{X}_1 | \mathbf{X}_2 | \mathbf{X}_3 | \mathbf{X}_4)$ which consists of four images, as shown in Fig.~\ref{fig:I2I_method}. $\mathbf{X}_i$,  $i \in {1, 2, 3, 4}$ are matrices, and  $|$ represents channel wise concatenation.
For each image, the pixel size is the same as the grid voxel size, defined earlier in this section. Accordingly, each of the images in Fig.~\ref{fig:I2I_method} is of dimension $ L \times W $. Thus, the dimension of $\mathbf{X}$ is $( L \times W \times 4)$. The first image, $\mathbf{X}_1$ has all pixel values $0$, except for the pixel containing the primary transmitter for the corresponding entry in $\mathcal{D}$, where the pixel value is $1$.
$\mathbf{X}_2$ has pixel value $1$ at locations where RSS measurements for the primary transmitter in $\mathbf{X}_1$ have been collected. However, we retain only $K$ measurements that are the nearest to the primary transmitter. This way, $\mathbf{f}$ learns to predict the RSS values in the vicinity of a transmitter, which is exactly what we want for the secondary transmitters during the online phase, as explained before. The third image, $\mathbf{X}_3$, is a map of the area, and the pixel values of $\mathbf{X}_3$ are between 0 and 1. Rather than using the actual map, we use a map obtained using RTI. The motivation for doing so was explained in Section~\ref{section:intro}. The details of creating the RTI-based map are described in Section~\ref{subsection:rti}. The fourth image, $\mathbf{X}_4$ captures the shadow fading on the links between the primary transmitter in $\mathbf{X}_1$ and the set of chosen measurement locations in $\mathbf{X}_2$. $\mathbf{X}_4$ is obtained by leveraging RTI and the details of creating $\mathbf{X}_4$ are also explained in Section~\ref{subsection:rti}. Finally, the output of $\mathbf{f}$ is an image where the pixels corresponding to the non-zero pixels in $\mathbf{X}_2$ represent the true RSS measured by the primary receivers considered in $\mathbf{X}_2$. The remaining pixels are not used for training; see~\eqref{eq:penalized_error}. 
These pixels are shown as $0$ in Fig.~\ref{fig:I2I_method}. 

For learning $\mathbf{f}$, we use the Adam optimizer~\cite{kingma2014adam} and minimize the following loss function using the training examples.
\vspace{-2mm}
\begin{equation} \label{eq:loss_func}
\vspace{-1mm}
    \mathcal{L} = \sum_{\mathcal{B}} \frac{\sum_{(i,j)} \pmb{\varepsilon}_b[i,j]}{\lambda_o \sum_{(i,j)}\mathbf{1}_{\mathbf{Y}_b[i,j] \leq \hat{\mathbf{Y}}_b[i,j]} + \lambda_u\sum_{(i,j)}\mathbf{1}_{\mathbf{Y}_b[i,j] > \hat{\mathbf{Y}}_b[i,j]}}
\end{equation} 
where $\mathcal{B}$ is a mini-batch of training examples, $b$ denotes an example belonging to $\mathcal{B}$, $\mathbf{1}_s$ is indicator function. $(\mathbf{X}_b, \mathbf{Y}_b)$ denotes the input-output pair of example $b$. $(i,j)$ runs over all pixels in $\mathbf{Y}_b$. $\pmb{\varepsilon}_b[i,j]$, the prediction error for pixel $(i,j)$ is:
\vspace{-2mm}
\begin{equation}\label{eq:penalized_error}
\vspace{-1mm}
    \pmb{\varepsilon}_b[i,j] = \begin{cases}
    0, \text{   if   }  \mathbf{Y}_b[i,j] = 0 \\
    \lambda_{o} \times \big|\mathbf{Y}_b[i,j] - \hat{\mathbf{Y}}_b[i,j]\big|, \text{   if   }   \mathbf{Y}_b[i,j] \leq \hat{\mathbf{Y}}_b[i,j]\\
    \lambda_{u} \times \big|\mathbf{Y}_b[i,j] - \hat{\mathbf{Y}}_b[i,j]\big|, \text{   if   }  \mathbf{Y}_b[i,j] > \hat{\mathbf{Y}}_b[i,j] \\
    \end{cases}
\end{equation}
where $\hat{\mathbf{Y}}_b[i,j]$ is the predicted RSS for pixel $(i,j)$. $\mathcal{L}$ represents the mean absolute error (MAE) loss function, but it penalizes the prediction error only for the pixels where the chosen $K$ receivers are present for example $b$. Additionally, we penalize underestimation errors more than overestimation errors. This is controlled by setting $\lambda_{u} > \lambda_{o}$, where $\lambda_{u}$ scales the underestimation errors and $\lambda_{o}$ scales the overestimation errors. 
Recall from Section~\ref{section:system_model} that the secondary transmitters will be activated based on the proactive spatial predictions by $\mathbf{f}$. Hence, as discussed in Section~\ref{section:intro}, we penalize $\mathcal{L}$ in the above manner so that the chances of activated secondary transmitters causing interference to primary receivers are reduced. However, caring only for the underestimation errors and ignoring the overestimation errors will reduce transmission opportunities for the secondary transmitters. Hence, $\lambda_{o}$ and $\lambda_{u}$ must be chosen carefully, and we do so via cross-validation. 

\begin{figure}[t]
    \centering
    \includegraphics[scale=0.4]{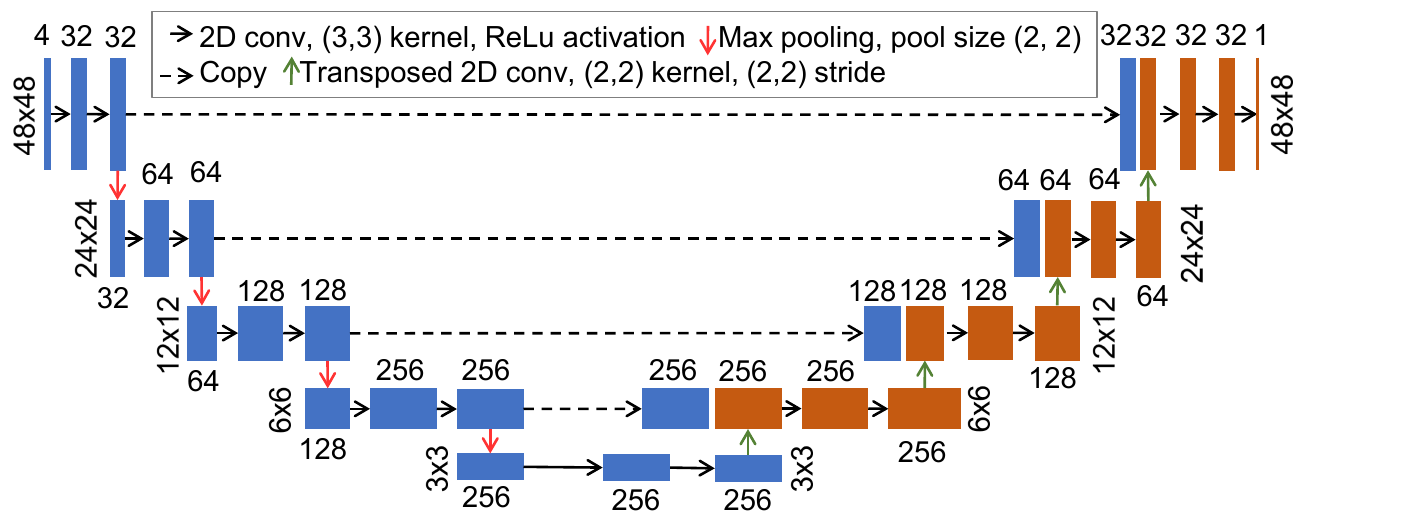}
    \caption{RSSu-net architecture. The numbers on the top of blocks are number of channels. The numbers in $a \times b$ format show the images' width and height.
    }
    \label{fig:unet_arcg}
\end{figure}

For our I2I formulation, we use the u-net NN architecture, shown in Fig.~\ref{fig:unet_arcg}, which is well suited for pixel-wise predictions. Alternative NN architectures are autoencoders, variational autoencoders, or any other architecture with an encoder-decoder structure. We experimented with some of these architectures but found that u-net performs better for our problem. The reasoning behind that is u-net combines (see `Copy' in Fig.~\ref{fig:unet_arcg}) 
the high-resolution features of the encoder/contracting path (see left half of Fig.~\ref{fig:unet_arcg}) and the outputs of transposed convolutions that increase the resolution of the output. Successive convolutions on the combined output lead to better predictions.
Since our u-net predicts RSS values, we call it RSSu-net. We use Keras~\cite{chollet2015keras} for the development of RSSu-net. Note that the input size of RSSu-net is determined by the grid voxel length $l$, which in turn depends on the density of the receivers that collect the RSS measurements. It is also evident from Fig.~\ref{fig:unet_arcg} that the NN's overall structure is governed by the input size. Hence, the density of collected measurements influences the learnability of $\mathbf{f}$. 

\textit{Inference}: The input to $\mathbf{f}$ during inference is $\mathbf{X}' = (\mathbf{X}_1^{'} | \mathbf{X}_2^{'} | \mathbf{X}_3^{'} | \mathbf{X}_4^{'})$. $\mathbf{X}'$ is similar to $\mathbf{X}$ with some differences.
The only non-zero pixel in $\mathbf{X}_1^{'}$ corresponds to the secondary transmitter's location for the corresponding input. $\mathbf{X}_2^{'}$ has pixel value $1$ at the $K$ chosen grid points for the secondary transmitter and $0$ otherwise. For a given secondary transmitter location, the $K$ chosen grid points are nearest to the transmitter but not under buildings/foliage. However, a subtle issue is knowing which grid points are valid (i.e., not under buildings/foliage), as a map is unavailable. In Section~\ref{subsection:rti}, we describe a method for extracting an RTI-based map, 
which can be used to extract valid grid points.
$\mathbf{X}_3^{'}$ is the same as $\mathbf{X}_3$ as it represents a map. It is the same for all the training and online examples. $\mathbf{X}_4^{'}$ captures the shadow fading on the links between the secondary transmitter in $\mathbf{X}_1^{'}$ and the set of grid points in $\mathbf{X}_2^{'}$. The output of $\mathbf{f}$ is an image where the pixels corresponding to the non-zero pixels in $\mathbf{X}_2^{'}$ represent the predicted RSS at the $K$ chosen grid points. The remaining pixels are ignored. 

\subsection{Assisting RSSu-net with RTI} \label{subsection:rti}
To deal with the challenge of the unavailability of area maps (buildings' locations and heights), we propose to use an RTI-based map. Specifically, using $\mathcal{D}$ we find the pixelated spatial loss field (SLF) vector, $\mathbf{p}$, of the area, scale its values between 0 and 1, reshape it to the shape of $\mathbf{X}_1$ and use it as $\mathbf{X}_3$. $\mathbf{p}$ is a vector that captures the SLF for each pixel in our considered area. We use the pixel size of SLF to be the same as our grid voxels defined in Section~\ref{subsection:unet}. Thus, $\mathbf{p}$ is a vector of length $ LW$. The interpretation of $\mathbf{p}$ is that the SLF value for a pixel captures its contribution to the shadow fading for any link that intersects this pixel~\cite{patwari2008effects}. The value of $\mathbf{p}$ for a particular pixel can be thought of as the amount of blockage that the pixel creates in RF sense. Hence, $\mathbf{p}$ can be used as a proxy for a map. The details of generating $\mathbf{p}$ and $\mathbf{X}_3$ are described below.

Using $\mathcal{D}$, we first form a shadow fading vector, $\mathbf{v}$.
\vspace{-2mm}
\begin{equation} \label{eq:shadow_eqn}
\vspace{-2mm}
    \mathbf{v} = \mathbf{z}_I - \mathbf{z}_M
\end{equation}
The elements of $\mathbf{v}$ correspond to the links used in the training of $\mathbf{f}$. Thus, the length of $\mathbf{v}$ is $KT_{PN}$. $\mathbf{z}_I$ is the ideal RSS vector based on the PLM, and $\mathbf{z}_M$ is the vector of measured RSS values on the links. 
The elements of $\mathbf{z}_I$, $\mathbf{z}_M$ are in dBm, and that of $\mathbf{v}$ in dB. For computing $\mathbf{z}_I$ for the $i^{th}$ link, we use 
\vspace{-2mm}
\begin{equation} \label{eq:plm}
\vspace{-2mm}
    \mathbf{z}_I[i] = z_0 - 10 \eta \log (d_i/d_0)
\end{equation}
where $\eta$ is radio wave propagation path loss exponent, $d_i$ is the distance of link $i$, $d_0$ is the smallest link distance among all links used in~\eqref{eq:shadow_eqn}, $z_0$ is the measured RSS on the smallest link. $\eta$ is estimated by fitting the collected data to~\eqref{eq:plm} using linear regression. Essentially,~\eqref{eq:shadow_eqn} models the difference between PLM-based RSS and measured RSS as shadow fading.

In RTI, the elements of $\mathbf{v}$ are modeled as a linear combination of the elements of the SLF vector $\mathbf{p}$:
\vspace{-2mm}
\begin{equation} \label{eq:rti_model}
\vspace{-2mm}
    \mathbf{v} = \mathbf{W}\mathbf{p} + \mathbf{n}
\end{equation}
where $\mathbf{n}$ is the measurement noise vector. $\mathbf{W}$ is a weight matrix that specifies the contributions of the elements
of $\mathbf{p}$ to the links. A common way of constructing $\mathbf{W}$ is the following.
\vspace{-2mm}
\begin{equation*} \label{eq:ellipse}
\vspace{-2mm}
    \mathbf{W}[k, q] = \frac{1}{\sqrt{d_{k}}} \begin{cases}
    1, \text{  if  } ||k_t - q_{c}|| + ||k_r - q_{c}|| \leq d_{k} + \lambda
    \\
    0, \text{  otherwise  }
    \end{cases}
\end{equation*}
where $k$ and $q$ are integers representing the row and column indices of $\mathbf{W}$, respectively. Different rows of $\mathbf{W}$ correspond to different links, and different columns of $\mathbf{W}$ correspond to different elements of $\mathbf{p}$. Although $\mathbf{p}$ is a vector, recall that the elements of $\mathbf{p}$ have a one-to-one association with the pixels in the rectangular grid of the area. $q_{c}$ is the coordinate of the center of the pixel that is associated with the $q^{th}$ element of $\mathbf{p}$. $d_{k}$ is the distance of link $k$. $k_t$ and $k_r$ are the coordinates of the transmitter and receiver of link $k$, respectively. Essentially the above equation says if we consider $k_t$ and $k_r$ as the foci of a narrow ellipse, then the elements of $\mathbf{p}$ whose associated pixels are inside the ellipse contribute towards $\mathbf{v}[k]$. The parameter $\lambda$ controls the ellipse width. Following the approach in~\cite{zhao2013radio}, using a regularized least square approach, we estimate $\mathbf{p}$ as:
\vspace{-2mm}
\begin{equation} \label{eq:least_square_rti}
\vspace{-2mm}
    \hat{\mathbf{p}} = \big(\mathbf{W}^{T}\mathbf{W} + \sigma_n^2 \mathbf{C}^{-1}\big)^{-1} \mathbf{W}^{T} \mathbf{v} 
\end{equation}
where $\sigma_n^2$ is the noise variance and $\mathbf{C}$ is the covariance matrix of the SLF. $\mathbf{C}[i,j]$, the covariance of the SLF between pixel $i$ and $j$ is modeled as, $\mathbf{C}[i,j] = (\sigma^2 / \delta) e^{-d_{ij}/\delta}$ where $i$ and $j$ are integers corresponding to row and column indices of $\mathbf{C}$, respectively. $\sigma^2$ is the variance of shadow fading between pixel $i$ and $j$, $d_{ij}$ is the distance between the centers of pixel $i$ and $j$, $\delta$ is a space constant. Once we obtain $\hat{\mathbf{p}}$, we scale it as $\hat{\mathbf{p}} = (\hat{\mathbf{p}} - \min(\hat{\mathbf{p}})) / (\max(\hat{\mathbf{p}}) - \min(\hat{\mathbf{p}}))$, reshape it to the shape of $\mathbf{X}_1$, and form a matrix. Then, we subtract all the matrix elements from 1 so that pixels with a higher SLF get a lower numerical value. The resulting matrix is used as $\mathbf{X}_3$.

In RSSu-net, we also assist our \NN using the estimated SLF vector, $\hat{\mathbf{p}}$, as described below. For an input to our NN, say $\mathbf{X}$, we compute the shadow fading 
for each of the links in that example as $\hat{\mathbf{v}} = \mathbf{W}\hat{\mathbf{p}}$. Then we scale $\hat{\mathbf{v}}$ as $\hat{\mathbf{v}} = (\hat{\mathbf{v}} - \min(\mathbf{v})) / (\max(\mathbf{v}) - \min(\mathbf{v}))$. 
Note that the minimum and maximum values of shadow fading are computed over 
$\mathbf{v}$ (refer to~\eqref{eq:shadow_eqn}) so that the scaling of $\hat{\mathbf{v}}$ is consistent across training and testing examples.
Next, we subtract the values of scaled $\hat{\mathbf{v}}$ from 1. Using the values after subtraction, we form a matrix $\pmb{\gamma}$ that has the same shape as $\mathbf{X}_1$ and non-zeros values for the pixels that are non-zero in $\mathbf{X}_2$. $\hat{\mathbf{v}}$ represents the shadow fading of the links between the transmitter in $\mathbf{X}_1$ and the receivers in $\mathbf{X}_2$. Hence, the values of $\hat{\mathbf{v}}$ (after scaling and subtracting from 1) are inserted in the pixels of $\pmb{\gamma}$ that correspond to the receivers' locations whose shadow fading is captured by $\hat{\mathbf{v}}$.
Finally, we use $\pmb{\gamma}$ as $\mathbf{X}_4$, as shown in Fig.~\ref{fig:I2I_method}. In Section~\ref{section:evaluations}, we show the benefits of appending $\mathbf{X}_4$ to the input of RSSu-net. In \sys, we adapt the version of RTI proposed in~\cite{wilson2010radio}. There are several variations of RTI. We use the vanilla version of RTI to show that RTI-based maps can be leveraged (account for unknown obstacles not on the map, deal with the unavailability of the map, and assist \DL model) in our problem. 
Integration of advanced RTI in \sys will be investigated in the future.

\subsection{Data Augmentation} \label{subsection:data_aug}
Recall from Section~\ref{section:system_model}, the number of examples in our training dataset is the same as the number of primary transmitters with unique locations in the crowdsourced data. However, the number of primary transmitters, $T_{PN}$, can be limited. In that case, the size of our training dataset will be small, leading to underfitting of our \NN. We propose a simple but effective data augmentation approach described below to avoid this problem.

As mentioned in Section~\ref{subsection:unet}, for each primary transmitter, we use $K$ unique RSS measurements (captured at $K$ unique receiver locations) near the transmitter. To increase the size of our training dataset, first, for each primary transmitter, we randomly sample the set of $K$ RSS measurements to extract $M; M > 1$ smaller subsets of RSS measurements of size $K/S; S > 1; \lfloor K/S \rfloor > 1$. The $M$ smaller subsets can have some common RSS measurements. Then, for each primary transmitter, we construct $M$ different training examples using the $M$ different subsets of RSS measurements. For these $M$ training examples, the matrix $\mathbf{X}_1$ (refer to Fig.~\ref{fig:I2I_method}) will be the same as the primary transmitter location is the same for all of them. However, the matrices $\mathbf{X}_2$ and $\mathbf{X}_4$ will differ for these $M$ training examples due to the subset sampling of the RSS measurements. Accordingly, the output images for these $M$ training examples will also differ. Using the above procedure, we can increase the size of our training dataset by $M$ fold. The parameters $M$ and $S$ ($K/S$ is the size of each subset) must be chosen carefully. If $S$ is small, then the different subsets of RSS measurements will overlap significantly, and we should use a smaller value of $M$ to avoid redundancy in the training examples. However, the benefit of the data augmentation will be minimal. In contrast, if $S$ is large, such that $K/S$ is small, the different subsets will have a low number of RSS measurements. 
In this case, we can have a large $M$, but each of the training examples will be too sparse for the \NN to learn effectively. As shown in Section~\ref{section:evaluations}, this data augmentation can significantly improve performance in RSSu-net.

\subsection{Boundary Proposal} \label{subsection:boundary_estimation}
After proactive RSS prediction for secondary transmitters, we draw boundaries around them and predict the out-of-zone power leakage. This must be done using predicted RSS values, i.e., before activating the secondary transmitters. Our boundary proposal method, an iterative search, is explained next.

For a secondary transmitter located at $(u',v')$, after the RSS prediction at the $K$ queried locations, we search for a set of grid points, $\mathcal{N}$ ($|\mathcal{N}| = N$, $N < K$, $\mathcal{N} \in \mathcal{K}$), where the predicted RSS values are below $z_{th} = N_f$ dBm. If the number of such grid points is less than $N$, we begin the second iteration in our search procedure where we look for $N$ grid points where the predicted RSS is below $z_{th} = N_f+g$ dBm, where $g > 0$ is a fixed parameter chosen by the SA. $g$ is a step size in our iterative search procedure that governs the increases in $z_{th}$ with every new iteration. We continue the search process until we find $N$ grid points. When the search procedure stops, the selected $N$ points become the proposed boundary for the secondary transmitter at $(u',v')$. If our search procedure stops after $m$ iterations for a particular secondary transmitter, then the out-of-zone power leakage of the proposed boundary is $z_{ooz} = N_f + (m - 1) \times g$ dBm. 
To ensure that the search procedure does not loop infinitely, we stop when $z_{th} \geq z_0$ (see~\eqref{eq:plm} for the definition of $z_0$). In such cases, the secondary transmitters is not allowed to transmit.

The boundary proposal algorithm only provides the boundary points. It does not pay attention to the points' connectivity and spread. Starting with the $K$ nearest grid points around a secondary transmitter as the query points addresses this problem to some extent by limiting the search region. This limitation of boundary proposal algorithm will be addressed in future. In our evaluations, we choose $N$ such that the boundary points for secondary transmitters form polygons around them.

\textit{Transmit Power Adaptation:} The final step in \sys is to adjust the transmit power of a secondary transmitter (refer to Fig.~\ref{fig:system_model_testing}). For a given set of interference protection boundaries for the primary transmitters and the proposed boundary for a secondary transmitter, the power adaptation is equivalent to the scaling $P_{PN}$ to $P_{SN}$ such that the secondary transmitter's proposed boundary and the primary transmitters' interference protection boundaries do not overlap. This scaling is dependent on estimating the interference protection boundaries for the primary transmitters, which is not the focus of our work. Hence, we do not explicitly evaluate the transmit power adaptation. Instead, we focus on the attributes of proposed boundaries for the secondary transmitters in our evaluations.

\section{Evaluations} \label{section:evaluations}
\begin{figure*}[t]
    \centering
    \begin{subfigure}[RTI not used]
    {\label{fig:no_rti_error}
    \includegraphics[scale=0.138]{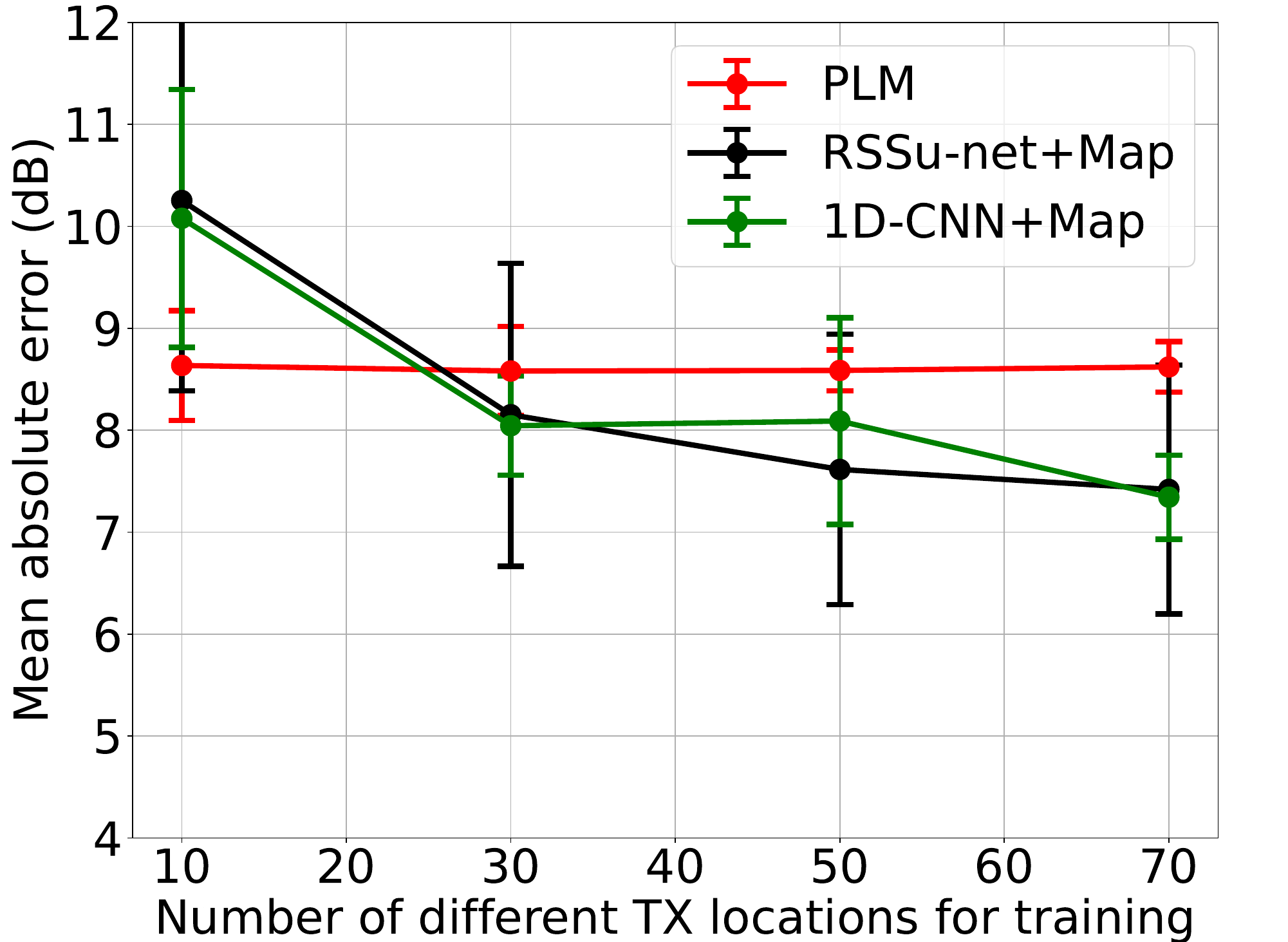}}
    \end{subfigure} 
    \hspace{-7mm}
    \begin{subfigure}[RTI is used, except PLM]
    {\label{fig:with_rti_error}
    \includegraphics[scale=0.135]{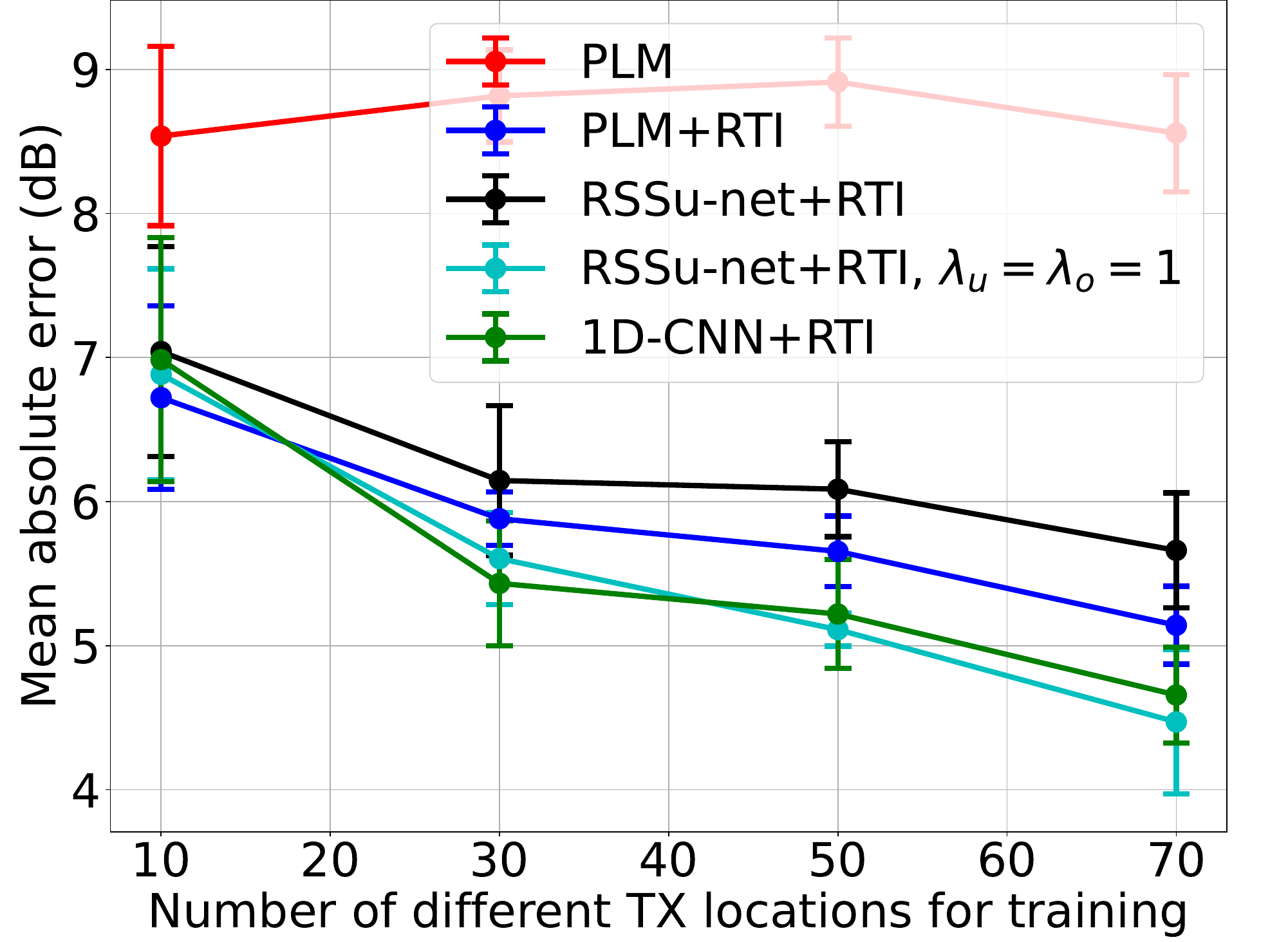}}
    \end{subfigure} 
    \hspace{-6mm}
    \begin{subfigure}[$\bar{|\epsilon|}$ using true RSS $>$ -50 dBm ]
    {\label{fig:with_rti_error_thresholded}
    \includegraphics[scale=0.135]{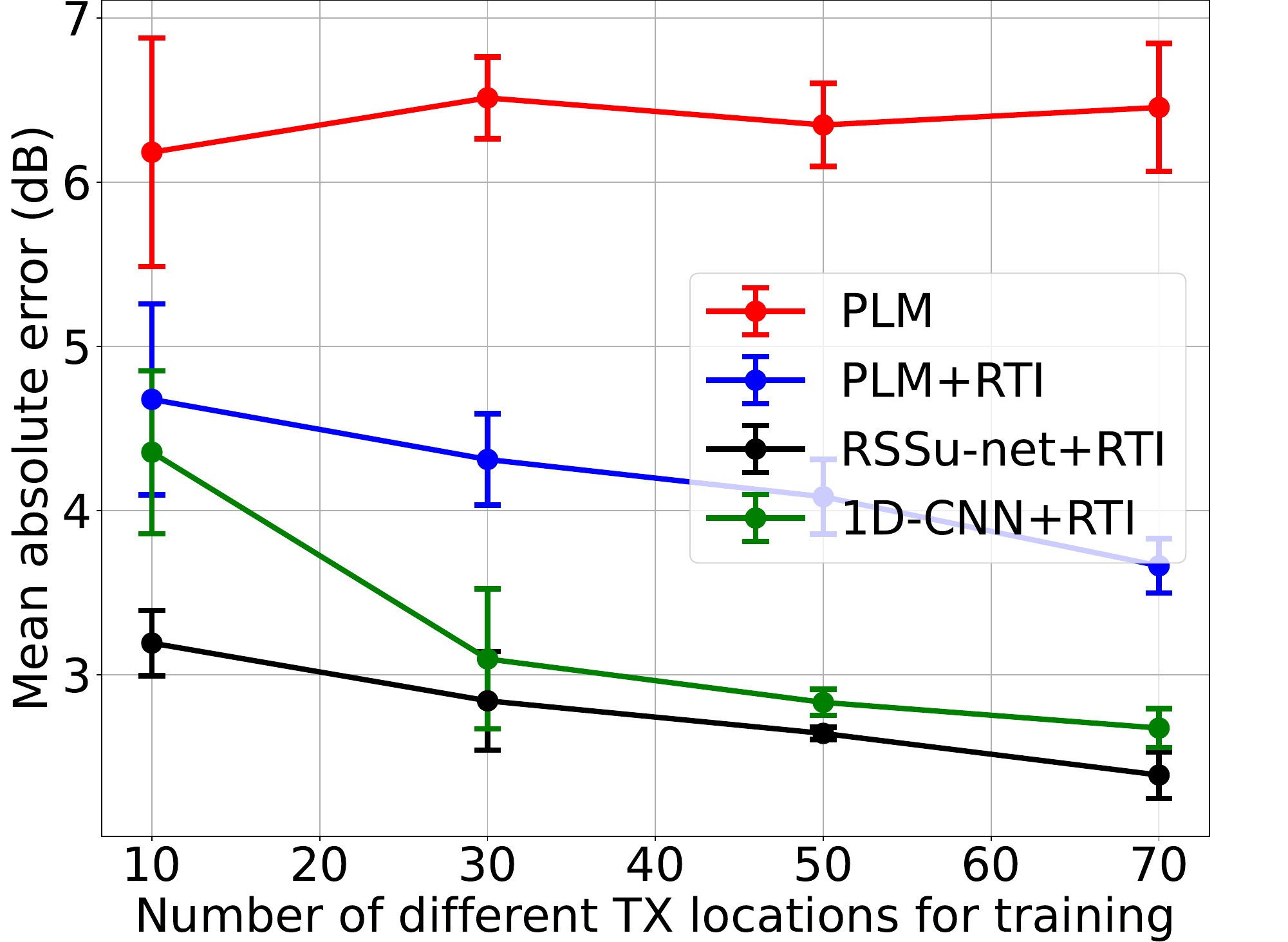}}
    \end{subfigure} 
    \hspace{-6.5mm}
    \begin{subfigure}[RTI is used for all except PLM]
    {\label{fig:with_rti_acc}
    \includegraphics[scale=0.139]{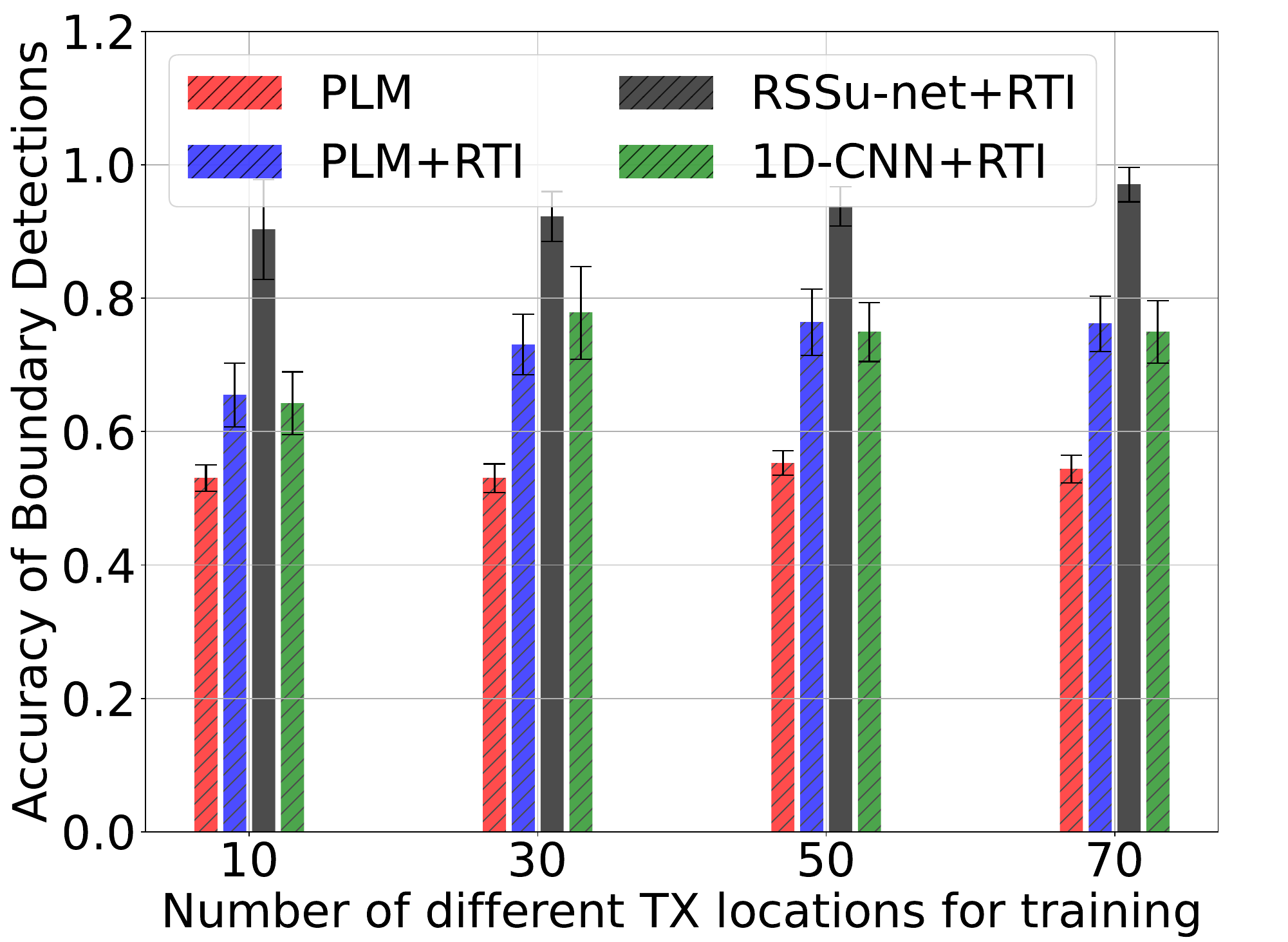}}
    \end{subfigure}
    \caption{Comparison of RSSu-net  with other methods in terms of $\bar{|\epsilon|}$ and $p_d$. Note the difference in the range of the Y-axis for different figures.} 
    \label{fig:main_result}
\end{figure*}

\textit{Data Collection}: We use the Wireless Insite~\cite{wirelessInsite} ray tracing software for generating our dataset.  We consider a 500 m $\times$ 500 m area with 15 square buildings of height 30 m and length of 20 m. Refer to $\mathbf{X}_3$ in Fig.~\ref{fig:I2I_method}, which is the map obtained from RTI, to get an idea of how the buildings are placed. In that image, the darker regions represent the buildings. We collect the dataset with 123 transmitters that transmit tone signals with a power of 30 dBm at a carrier frequency of 1 GHz and effective bandwidth of $B=1$ MHz. For each transmitter, the receivers are placed over the whole area on a grid with a grid voxel length of 10 m, but we consider only the outdoor ones.

\textit{Training and Test Data}: Among the 123 transmitters in the generated dataset, we use at most 70 (sometimes less than 70 to evaluate the impact of $T_{PN}$ on various aspects of \sys) as primary transmitters that become our training transmitters. Among the remaining ones, we randomly select 30 as secondary transmitters for testing. The receivers for the training transmitters and the query points for the test transmitters are selected as described in Section~\ref{subsection:unet}. Each of the data points in our results is based on 5 iterations. Each iteration consists of training and testing with randomly selected transmitters and receivers. We use multiple iterations 
so that our observations are not biased on a particular layout of transmitters and receivers.

\textit{\sys Parameters}: In our evaluation of \sys, we use grid voxel length, $l=10$ m and $K=$ 200. 
For the data augmentation, we use $S = 5$ and $M= 200$. For the boundary proposal, we use $N = 20$ and $g = 10$ dB. Finally, in our RTI approach, we use $\lambda = 5$ m, $\sigma_n^2 = 1$, $\sigma^2 = 0.5$, and $\delta = 1$.

\begin{figure}[t]
    \centering
    \includegraphics[scale=0.37]{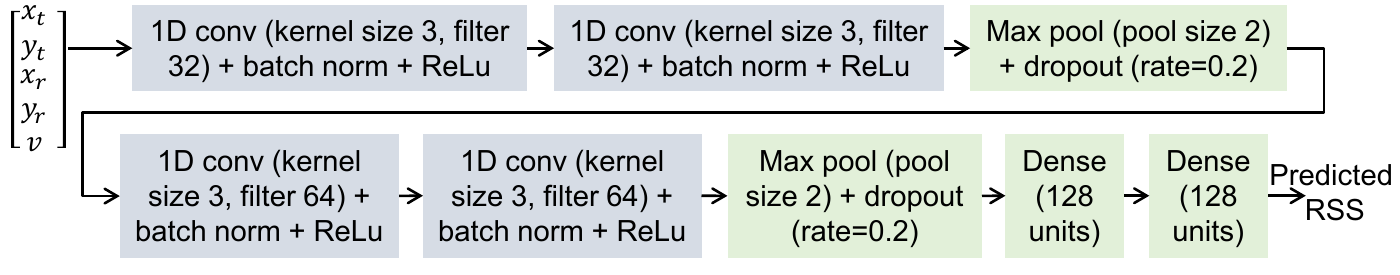}
    \caption{NN architecture for 1D-CNN+RTI.
    }
    \label{fig:1dcnn_arch}
\end{figure}

\textit{Methods for Comparison}: We compare the performance of the following proactive spatial prediction methods with RSSu-net. While evaluating these methods, we still rely on our \sys framework. Thus, we compare these methods with RSSu-net within the framework of \sys. We consider these methods for comparison as we believe these represent the different learning and non-learning proactive spatial prediction methods that are applicable in the framework of \sys.
\begin{itemize}
    \item \textit{PLM}: This method uses~\eqref{eq:plm} to predict RSS at the $K$ nearest locations for each secondary transmitter considered in the testing phase. PLM relies on the training data only to find the $\eta$ and $z_0$ in~\eqref{eq:plm} using linear regression.
    \item \textit{PLM+RTI}: This method is similar to PLM, but we add the shadow fading for each of the links for which RSS is to be predicted. The shadow fading is obtained using the RTI method described in Section~\ref{subsection:rti}. PLM+RTI relies on training data to estimate the SLF, $\hat{\mathbf{p}}$ (see Section~\ref{subsection:rti}).
    \item \textit{1D-CNN+RTI}: We develop this method as another baseline where we use a 1D CNN for link-wise RSS prediction. The 1D CNN takes the coordinates of the transmitter and receiver of a link and predicts the RSS at the receiver. It also uses the shadow fading loss of the link as input. For example, if the transmitter is at $(x_t, y_t)$ and the receiver at $(x_r, y_r)$, and the shadow fading on the link (as computed in Section~\ref{subsection:rti}) is $v$, then the input to the \NN is $[x_t, y_t, x_r, y_r, v]^T$. Each of the values in this input vector is scaled to lie between 0 and 1 based on its maximum and minimum values, as was done for $\hat{\mathbf{p}}$ in Section~\ref{subsection:rti}. As the number of links in our training dataset is large, this method does not require data augmentation. The \NN architecture used for this method is shown in Fig.~\ref{fig:1dcnn_arch}.
    \item \textit{1D-CNN:}
    This method is similar to 1D-CNN+RTI, but RTI is not used. Hence, the link shadow fading, $v$, is not used as input to the \NN.
    \item \textit{RSSu-net+Map:} This method is similar to RSSu-net, but RTI is not used. Hence, $\mathbf{X}_4$ (refer to Fig.~\ref{fig:I2I_method}) is not used, and for $\mathbf{X}_3$ the actual map of the area is used. For using the actual map as $\mathbf{X}_3$, we scale the height of every pixel to lie between 0 and 1, and subtract these values from 1 as done for $\hat{\mathbf{p}}$ in Section~\ref{subsection:rti}.
\end{itemize}
\textit{Metrics}: We use the following metrics for our evaluation.
\begin{itemize}
    \item $\bar{|\epsilon|}$ (dB): Mean absolute error of RSS predictions for the secondary transmitters. The mean is computed over the $K$ chosen grid points for each test secondary transmitter.
    \item $p_d$: Mean accuracy of our boundary proposal algorithm, where accuracy is defined as follows. For a secondary, if the RSS prediction for a boundary point is overestimated, then we call it an accurate prediction. A higher value of $p_d$ implies a lower probability of interference to the coexisting primary receivers. The mean is computed over the $N$ boundary points for each test secondary transmitter.
    \item $\bar{z}_{ooz}$ (dBm): Mean of out-of-zone power leakage, $z_{ooz}$, of proposed boundaries for the secondary transmitters. The mean is computed over different secondary transmitters.
\end{itemize}
\begin{figure}[t]
    \centering
\includegraphics[scale=0.295]{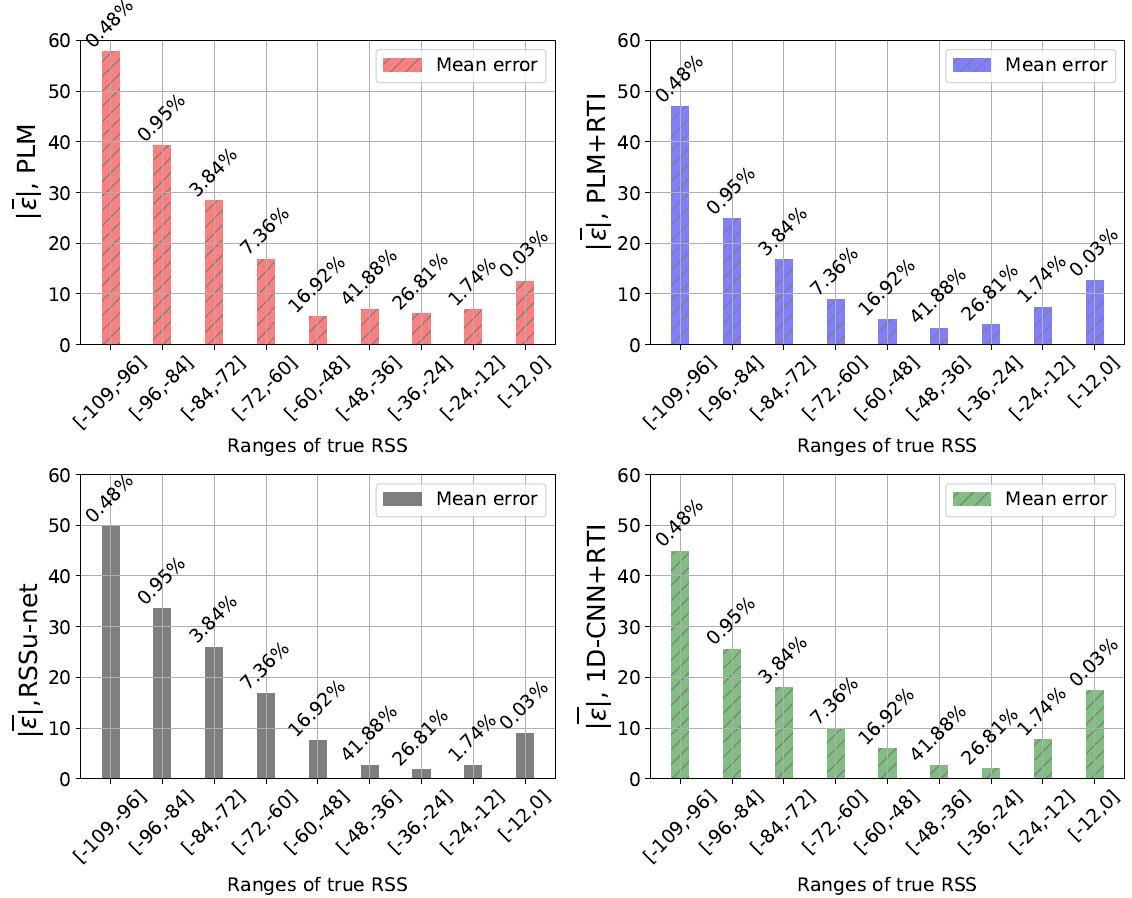}
    \caption{$\bar{|\epsilon|}$ of different methods for different ranges of true RSS (in dBm).} 
    \label{fig:error_histogram}
\end{figure}

\subsection{Results}
\textit{Impact of RTI:} We show the impact of RTI in \sys using Fig.~\ref{fig:no_rti_error} and~\ref{fig:with_rti_error}. Fig.~\ref{fig:no_rti_error} shows $\bar{|\epsilon|}$ of different methods for different values of $T_{PN}$. In Fig.~\ref{fig:no_rti_error}, RTI is not used; hence, RTI-based methods are not shown. We observe that with a higher value of $T_{PN}$, $\bar{|\epsilon|}$ for RSSu-net+Map reduces. With $T_{PN} = 70$, RSSu-net+Map is better than PLM by 2 dB. We also see that 
1D-CNN performs comparably to RSSu-net+Map. These observations show that learning-based proactive spatial predictions are more accurate than model-based predictions. Fig.~\ref{fig:with_rti_error} is similar to~\ref{fig:no_rti_error}, but in this case, we use RTI. We see that using RTI improves RSSu-net's performance by 2 dB further. This shows the merit of assisting our \DL model via RTI.

\textit{Average RSS prediction error:} Fig.~\ref{fig:with_rti_error} shows that both 1D-CNN+RTI and PLM+RTI perform slightly better than our proposed method. This performance gap can be attributed to the fact that unlike RSSu-net, 1D-CNN+RTI and PLM+RTI do not penalize overestimation and underestimation errors differently. To justify this argument, we also plot a curve labeled `RSSu-net+RTI, $\lambda_u = \lambda_o = 1$', which is the same as RSSu-net but uses $\lambda_u = \lambda_o = 1$ in~\eqref{eq:penalized_error} during its training, i.e., it does not penalize overestimation and underestimation errors differently and aims to minimize the MAE. Fig.~\ref{fig:with_rti_error} shows that `RSSu-net+RTI, $\lambda_u = \lambda_o = 1$' is better than both 1D-CNN+RTI and PLM+RTI, for higher values of $T_{PN}$. In summary, in terms of mean error, our proposed method is comparable to other methods that rely on RTI. RSSu-net does not achieve the lowest $\bar{|\epsilon|}$ as it has not been trained to do so.

\begin{table}[b]
\caption{Out-of-zone power leakage of proposed boundary}
\label{tab:boundary}
\centering{
{
\begin{tabular}{ c |c |c| c}
\hline\hline
$T_{PN}$ & RSSu-net & 1D-CNN+RTI & PLM+RTI\\
 & $\bar{z}_{ooz}$ (dBm) & $\bar{z}_{ooz}$ (dBm) & $\bar{z}_{ooz}$ (dBm)\\\hline
70 & $-45$ & $-53$ & $-53$ \\
\hline
\end{tabular}}}
\end{table}

\textit{Variation of $\bar{|\epsilon|}$ with target RSS:} Fig.~\ref{fig:error_histogram} shows $\bar{|\epsilon|}$ for different methods for different ranges of target/true RSS values. The numbers on the top of the bars show the percentage of test data that falls in the target RSS range corresponding to that bar. The figure shows that all the methods perform poorly when the target RSS, which must be estimated, is very low. The low target RSS values represent scenarios that cannot be well captured by radio wave propagation models. Hence, the model-based approaches, PLM and PLM+RTI, suffer in the low target RSS regime. The \DL approaches have poor performance in this regime because the low RSS measurements are infrequent in the training data.
The DL models do not get enough training data in the low RSS regime to learn well about such scenarios. Although the percentages in Fig.~\ref{fig:error_histogram} are for test data, the percentages in training data follow a similar trend.

To eliminate the impact of the low target RSS values in $\bar{|\epsilon|}$, in Fig.~\ref{fig:with_rti_error_thresholded}, we compare $\bar{|\epsilon|}$ for the different methods, but the errors are considered only if the true RSS is above -50 dBm. We observe that $\bar{|\epsilon|}$ for all methods have reduced compared to Fig~\ref{fig:with_rti_error}. This is expected as the high errors for the low RSS values are not considered in Fig.~\ref{fig:with_rti_error_thresholded}. Interestingly, in Fig.~\ref{fig:with_rti_error_thresholded}, RSSu-net is the best-performing method. This can be explained using Fig.~\ref{fig:error_histogram}, where we see that for higher values of true RSS, RSSu-net is better than other methods.
\begin{figure}[t]
    \centering
    \includegraphics[scale=0.38]{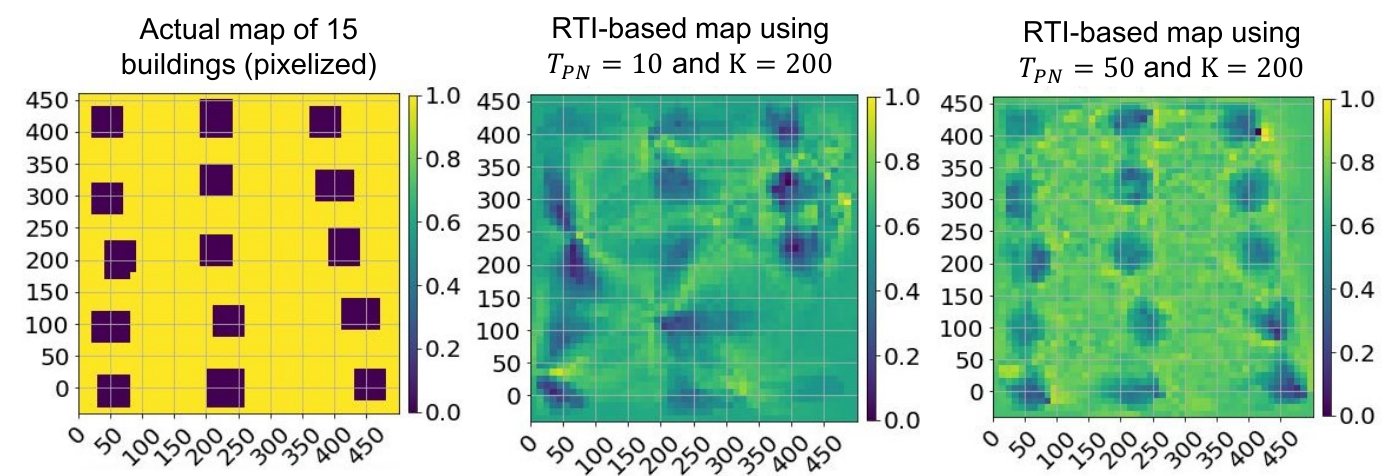}
    \caption{Figure shows the impact of $T_{PN}$ on the quality of RTI-based map.}
    \label{fig:rti_quality}
\end{figure}

\textit{Quality of RTI-based map:} Fig.~\ref{fig:rti_quality} shows the impact of $T_{PN}$ on RTI-based maps. It shows, with $T_{PN}= 50$, the RTI-based map resembles the actual map closely. However, a low value of $T_{PN} = 10$ leads to an inferior quality of the RTI-based map. This happens because the RTI-based map is estimated using the linear model of~\eqref{eq:rti_model}. The estimation improves with the number of equations in~\eqref{eq:rti_model}, which is dictated by $T_{PN}$.

\begin{table}[b]
\caption{Miscellaneous evaluations for $T_{PN} = 70$}
\label{tab:misc}
\centering{
\resizebox{3.5in}{!}
{
\begin{tabular}{ c |c  c |c  c| c c}
\hline\hline
Setup & \multicolumn{2}{|c}{RSSu-net}& \multicolumn{2}{|c}{1D-CNN+RTI} & \multicolumn{2}{|c}{PLM+RTI}\\
 & $\bar{|\epsilon|}$ & $p_d$ & $\bar{|\epsilon|}$ & $p_d$  & $\bar{|\epsilon|}$ & $p_d$ \\\hline
8 buildings  & $5.4$ & $0.98$ & $3.1$ & $0.88$ & $3.5$ & $0.86$ \\
15 buildings+foliage & $6.9$ & $0.87$ & $6.8$ & $0.77$ & $6.0$ & $0.73$\\
RSS, location errors &$ 5.7$ & $0.85$ & $4.6$ & $0.7$ & $5.3$ & $0.7$\\
No augmentation &$ 7.8$ & $0.97$ & NA & NA & NA & NA\\
\hline
\end{tabular}}}
\end{table}

\textit{Accuracy of boundary proposal:} Our goal in \sys is to achieve a higher boundary proposal accuracy, $p_d$, while having reasonable performance in terms of $\bar{|\epsilon|}$. Fig.~\ref{fig:with_rti_acc} shows the performance of different methods in terms of $p_d$ for different values of $T_{PN}$. For this plot, we use the boundary proposal algorithm described in Section~\ref{subsection:boundary_estimation}, but vary the methods for proactive RSS prediction. We observe that RSSu-net performs better than all other methods, and $p_d$ improves, in general, with a higher number of $T_{PN}$. For $T_{PN} = 70$, RSSu-net achieves $p_d = 0.97$, which is at least 19\% better than the $p_d$ with any other method. RSSu-net achieves this performance improvement due to our loss function in~\eqref{eq:loss_func} and carefully chosen hyperparameters in~\eqref{eq:penalized_error}.

\textit{Out-of-zone power leakage:} Next, we show the out-of-zone power leakage of the boundary proposal method in Table~\ref{tab:boundary} for different proactive RSS prediction methods with $T_{PN} = 70$. We see that the $\bar{z}_{ooz}$ for the different methods is comparable, but RSSu-net has 8 dB higher $\bar{z}_{ooz}$ than that of others. This happens because RSSu-net has been trained to reduce the underestimation errors at the cost of relatively higher overestimation errors. Hence, it overestimates the out-of-zone power leakage of the proposed boundaries. 

In our setup, $N_f = -100$ dBm, but the value of $\bar{z}_{ooz}$ for all methods is much higher than that. This observation can be explained using Fig.~\ref{fig:error_histogram}, which shows that the percentage of target RSS measurements in the low RSS regime is very low. Additionally, the estimation errors in the low target RSS ranges are primarily overestimation errors (not shown in Fig.~\ref{fig:error_histogram} due to space constraints). These two factors collectively make the possibility of capturing $N = 20$ points with a very small RSS low. A relatively higher estimated ${z}_{ooz}$ than actual ${z}_{ooz}$ can lead to unnecessary transmit power reduction for the secondary transmitters, i.e., subpar usage of the shared spectrum. However, this issue will not negatively impact the primary receivers in terms of interference.
\begin{table}[b]
\caption{Average prediction time of RSSu-net versus ray tracing}
\label{tab:run_time}
\centering{
{
\begin{tabular}{ c |c }
\hline\hline
Ray tracing & RSSu-net\\\hline
$20 \times 10^3$ ms & 10 ms  \\
\hline
\end{tabular}}}
\end{table}

\textit{Miscellaneous evaluations}: In Table~\ref{tab:misc}, we show several important aspects of \sys that demonstrate its generality and robustness. All the results in Table~\ref{tab:misc} are for $T_{PN} = 70$.

In the first row of Table~\ref{tab:misc}, 
we show the impact of having a less complex RF environment. We consider a dataset similar to the one described before, but with 8 buildings instead of 15. By comparing this row with Fig.~\ref{fig:with_rti_error} and~\ref{fig:with_rti_acc}, we observe that a lesser complex RF environment leads to a lower value of $\bar{|\epsilon|}$ for all the methods.
Importantly, RSSu-net retains its superiority over other methods in $p_d$. In contrast to the first row, we make the RF environment more complex than our original dataset in the second row of Table~\ref{tab:misc}. Specifically, we consider another dataset similar to our original dataset (15 buildings) but with foliage in the area. By comparing this row with Fig.~\ref{fig:with_rti_error} and~\ref{fig:with_rti_acc}, we see that a complex RF environment increases $\bar{|\epsilon|}$ for all methods, but RSSu-net maintains the highest $p_d$. The summary of the first two rows of Table~\ref{tab:misc} is that data-driven models better capture simpler RF environments than complex ones when the amount of available data is the same. 

Next, we show the impact of measurement errors in the third row of Table~\ref{tab:misc}. For that, we consider our original dataset (15 buildings) but assume that the crowdsourced measurements from the primary receivers can have random location errors (5 m on average) and RSS measurement errors (5 dB on average). Comparing this row with Fig.~\ref{fig:with_rti_error} and~\ref{fig:with_rti_acc}, we see that while the $\bar{|\epsilon|}$ for different methods remains similar, the $p_d$ drops for all methods due to measurement errors. Again, RSSu-net has better $p_d$ than the other methods.

Using our original 15 buildings dataset, the last row of Table~\ref{tab:misc} shows that when data augmentation is not used, RSSu-net can have inferior performance, $\bar{|\epsilon|} = 7.8$, compared  to the case when data augmentation is used, $\bar{|\epsilon|} = 5.5$ in Fig.~\ref{fig:with_rti_error}. This happens because, without data augmentation, the amount of training data is insufficient for RSSu-net to learn effectively.

\textit{Prediction time:} Finally, in Table~\ref{tab:run_time}, we show the time needed for RSSu-net and ray tracing to predict the RSS at $K=200$ query points for a secondary transmitter. As discussed in Section~\ref{section:intro}, the prediction time for ray tracing is significantly higher than that of RSSu-net. Moreover, ray tracing requires the area map for making predictions.

\section{Conclusions and Future Work} \label{section:conclusions}
We developed a novel framework called \sys that can dynamically create boundaries around secondary transmitters without any active transmission from them. 
We proposed an I2I \DL method called RSSu-net that can perform the proactive spatial predictions for the secondary transmitters. We developed a boundary proposal algorithm that operates on proactive predictions and creates the boundaries around the secondary transmitters. We thoroughly evaluated \sys under different settings and showed that RSSu-net performs reasonably well in terms of the average prediction error of RSS values, $\approx$ 5 dB mean absolute error. Importantly, \sys creates proactive boundaries around the secondary transmitters such that these transmitters can be activated with $\approx$ 97\% probability of not interfering with the primary receivers.

In our current framework, we assumed all the transmitters to have omnidirectional antennas. In the future, we will investigate how \sys can be used for directional transmitters.

\section*{Acknowledgment}
This work was supported by NSF SpectrumX via award 2132700, NSF grant 2003098, and a gift from Intel.

\bibliographystyle{unsrt}
\bibliography{Main}

\end{document}